\documentclass[10pt,aps,prb,twocolumn,floatfix,preprintnumbers]{revtex4}
\usepackage{graphicx}
\usepackage{epsf}
\usepackage{hyperref}
\usepackage{amsmath}    % contains the advanced math extensions for LaTeX
\usepackage{amssymb}    % adds new symbols in to be used in math mode
\usepackage{xcolor}
\usepackage{svg}

\begin{document}
\title{{\em Ab initio} simulation of the structure and transport properties of zirconium and  ferromagnetic cobalt contacts on the two-dimensional semiconductor WS$_2$}
\author{Hamideh Kahnouji$^{1,2}$}
\author{Peter Kratzer$^{2}$}
\author{S. Javad Hashemifar$^{1}$} 
\affiliation{$^{1}$Department of Physics, Isfahan University of Technology, 84156-83111 Isfahan, Iran}
\affiliation{$^{2}$Faculty of Physics  and Center for Nanointegration (CENIDE) , University of Duisburg-Essen, Lotharstrasse 1, 47057 Duisburg, Germany}

\begin{abstract}
Using density-functional theory calculations, the atomic and electronic structure of single-layer WS$_2$ attached to Zr and Co contacts are determined. 
Both metals form stable interfaces that are promising as contacts for injection of n-type carriers into the conduction band of WS$_2$ with Schottky barriers of 0.45eV and 0.62eV for Zr and Co, respectively. With the help of quantum transport calculations, we address the conductive properties of a free-standing WS$_2$ sheet suspended between two Zr contacts. It is found that such a device behaves like a diode with steep I-V characteristics. Spin-polarized transport is calculated for such a device with a floating-gate Co electrode added. Depending on the geometrical shape of the Co gate and the energy of the carriers in WS$_2$, the transmission of spin majority and minority electrons may differ by up to an order of magnitude. Thus the steep I-V characteristics of the nanoscale device makes it possible to realize a spin filter. 
\end{abstract}

\maketitle
%%%%%%%%%%%%%%%%%%%%%%%%%%%%%%%%%%%%%%%%%%%%%%%%
\section{Introduction}

Transition metal dichalcogenide (TMDC) monolayers (MLs) 
have emerged as a promising 2D crystal family with several characteristic features. 
Their direct band gap and the strong 
spin-orbit coupling combined with the lack of inversion symmetry leads to a unique coupling of the spin and valley degrees of freedom. 
Due to this distinct feature, elastic scattering of charge carriers is possible 
only by simultaneously flipping the spins of two carriers, and this severe restriction leads to small inter-valley scattering rates and therefore
long spin and valley lifetimes. These properties make the TMDCs 
promising candidates for future nanoscale electronics. 
The role of contacts  of 2D materials is an important 
issue for electronics device performance.  Due to the semiconducting nature
of  2D TMDCs, metal contacts are likely to create Schottky barriers. On the one hand, these barriers 
can be useful in coping with the impedance mismatch problem \cite{Schmidt2000}, as has been pointed out in the field of spintronics\cite{Fert2001}.
On the other hand, an elimination of Schottky barriers is desirable if the injection current should be maximized. 

Several studies have focused on achieving desirable contact properties. Experimentally, the growth of metals on TMDCs have  affected the resulting morphologies; for example, 
Pd grows on MoS$_{2} $  as a uniform layer while Au and Ag grow in clusters\cite{doi:10.1021/nn4052138}.
Both experiments\cite{6724660}  and simulations~\cite{Kang2012ACS}  have predicted that Ti as
contact metal to MoS$_{2}$ helps to eliminate the Schottky barrier.
In 2016, Farmanbar and Brocks  calculated  the  contact  between
MoS$ _{2} $ and various metal surfaces, such as  Au, Ag, Cu, and
Ni.  Their results show  that MoS$_{2}$  interacts strongly with 
transition metals~\cite{PhysRevB.93.085304}. DFT calculations have shown that the tunnel barrier between Pd and monolayer MoS$ _{2}$  can be reduced to negligible height\cite{Kang2012ACS}. Kang {\it et al.} studied  the
interface  between monolayer MoS$_{2}$ and WSe$_{2}$ 
and various metals (including  In, Ti, Au, Pd, Mo, and W) by DFT calculations. Their results show that Ti and Mo are the best metals for n-type contacts  on monolayer  MoS$_{2}$,  while Pd was found to be  the best metal for monolayer  WSe$ _{2}$, both metals forming n-type contacts~\cite{PhysRevX.4.031005}. 

Most studies examining TMDCs for device applications have
reported on interfaces between TMDC MLs and various non-magnetic metals. 
Only few studies have been focused on interfaces between  ferromagnetic metals and TMDC MLs. Experimentally, 
magnetic tunnel junctions of Fe/MoS$_{2}$/Fe  have been
investigated  and the results demonstrated that MoS$ _{2}$  became conductive when the MoS$_{2}$  spacer contained only one or two layers~\cite{PhysRevB.90.041401}. Chen {\it et al.} have investigated  properties of  single-layer MoS$_{2}$  with Co electrodes and measured the Schottky barrier height in this system. They found  that by inserting a thin MgO oxide barrier
between the Co electrode and the MoS$_{2}$  flake, the Schottky
barrier height can be reduced~\cite{doi:10.1021/nl4010157}.
Finally, Garandel {\it et al.} have recently studied  the electronic structure of Co/MoS$_{2}$  interfaces and the Schottky barrier height of this  system has been estimated to be $\sim0.32$~eV~\cite{PhysRevB.95.075402}.

It is important to understand the nature of the electronic interface between metals and a TMDCs monolayer. Hence,
in this paper, we employ first principles computations 
to study the properties of the interface WS$_{2}$/metal.  
Slabs of the hexagonally close-packed (hcp) metals Zr and Co are used to model the (0001) surfaces of Zr bulk or Co bulk, which serve as templates to place the hexagonal layer of WS$_2$. 
WS$ _{2}$  has attracted appreciable interest among TMDC monolayers due to  the presence of a high valence band splitting and a high mobility. 
We considered Zr and Co as suitable candidates for contact properties. Zr emerges as an ideal candidate with only 1.5\% mismatch to
WS$ _{2} $. Co appears less suitable at first sight due to a large lattice mismatch, but spin-injection from such a high-Curie-temperature
ferromagnetic metal to a single TMDC layer is very important since it would give access to TMDC-based spintronic devices.

\section{Computational details}
All presented calculations are performed in the frame-
work of the spin-polarized Kohn-Sham density functional
theory by using the all-electron full-potential code FHI-aims\cite{FHI-aims}.  
The generalized gradient approximation (GGA) in the Perdew-Burke-Ernzerhof (PBE)\cite{PhysRevLett.77.3865}
implementation  is applied as the exchange-correlation functional. 
All calculations reported here are done in the scalar-relativistic approximation with the so-called 'tight'  settings for the basis set and on grids in the FHI-aims code.
A convergence criterion of $10^{-5}$eV for the total energy and $10^{-5}$e/{\AA}$^{3}$  for the charge density were employed. 
To treat van der Waals (vdW) interactions, we apply the Tkatchenko-Scheffler (TS) method\cite{PhysRevLett.102.073005}, i.e. a density-dependent pairwise interaction is added to the exchange-correlation energy to describe the interfaces Co/WS$_{2}$ and Zr/WS$ _{2}$.
For the Zr or Co metal electrodes, only the interface Zr or Co layer takes part in the vdW interaction, and modified vdW parameters according to the TS scheme for surfaces \cite{Ruiz12} are used. Van der Waals interactions between Co or Zr atoms having a bulk environment are disregarded.

Before constructing  supercells for calculations of WS$ _{2}$ on cobalt, first  the theoretical lattice constants in the hexagonal plane of hcp  bulk cobalt and of a WS$ _{2}$ ML were obtained to be   
$a_{\rm Co} = 2.49${\AA}  and $a_{\rm WS_2} = 3.18 ${\AA}, respectively, in satisfactory agreement with the experimental values of 2.49{\AA} and 3.08{\AA} \cite{PhysRevB.60.4397,2053-1583-4-1-015026,doi:10.1063/1.4774090}, respectively.  

For the Co metal electrode, we consider two possible orientations of the WS$_2$ layer. 
In the first orientation, termed Co/WS$ _{2}R(0^{\circ})$, the lattice vectors of the Co(0001) and the WS$_2$ unit cells are collinear. 
The Co/WS$_{2}R(0^{\circ})$  hybrid structure was modeled as a periodic slab by setting the supercell lattice parameter to $a=b=12.485${\AA}, corresponding to $5 \times 5$ primitive cobalt cells and $4 \times 4$ primitive WS$_{2}$ cells.   

As an alternative, we consider a supercell in which the misfit strain between Co and WS$_2$ is (almost) compensated only in one crystallographic direction, while there is a relatively large strain in the perpendicular direction. 
This situation, termed Co/WS$ _{2}R(90^{\circ})$, is realized if the WS$ _{2} $ monolayer is deposited with a rotation of 90$^{\circ}$ with respect
to the cobalt slab. 
We used a $\sqrt{3} \times 4$ supercell of WS$_{2}$ matched to a $2  \times 3\sqrt{3}$ supercell of the Co(0001) surface. 
In this situation, a pre-strain in  WS$ _{2} $ is introduced
that is 1.73\% for Co/WS$ _{2}R(0^{\circ})$ and 10.28\% at the Co/WS$ _{2}R(90^{\circ})$ interface.
.pdf

 \begin{figure}[ht]
 \includegraphics[scale=0.42]{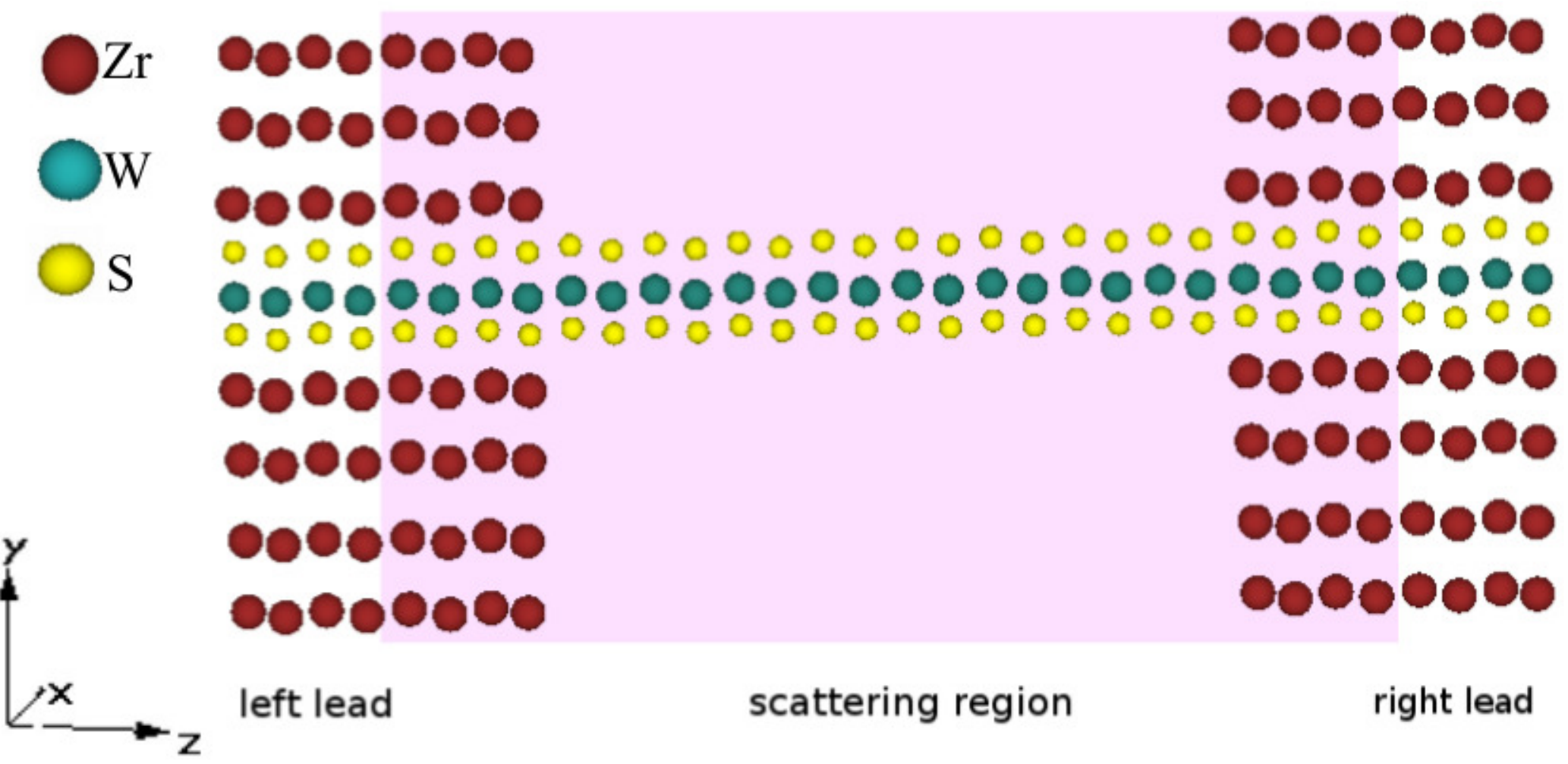}\\
 \caption{\label{fig:diode}
 Ballistic conductance setup, consisting of the left and right Zr leads and the scattering region (shaded). }
 \end{figure}

The calculated in-plane lattice constant of bulk hcp Zr is $a= 3.23${\AA}, in perfect match with the experimental value of 3.23{\AA} \cite{WILLAIME2003205}. 
To model the Zr/WS$ _{2}$ interface, the WS$_{2}(1\times 1)$ unit cell is adjusted to the  $(1  \times 1)$ unit cell of Zr(0001). 
In all calculations, the WS$_2$ layer was attached only to one side of the metal slab, and a 15{\AA} vacuum layer was added to prevent artificial interaction due to the periodic boundary conditions perpendicular to the surface. 
In the geometry optimizations, a convergence criterion of $10^{-2}$~eV/{\AA}  for
the residual forces was adopted.
The Brillouin zone integrations were performed by using  
 meshes of $10\times10\times1$ , $14\times10\times1$  and  $14\times14\times1$ {\bf k}-points for  Co/WS$ _{2}R(0^{\circ})$ ,  Co/WS$ _{2}R(90^{\circ})$ and Zr/WS$ _{2}$, respectively.
Once the atomic structure has been obtained, we  include spin-orbit coupling effects on the
electronic structure via the second-order variational method implemented in FHI-aims \cite{Huhn17}.

For the simulation of the transport characteristics of spintronics devices based on a WS$_2$ sheet, we 
used the PWCOND code\cite{PhysRevB.59.2267,PhysRevB.70.045417}  of the {\sc Quantum Espresso} package\cite{PWSCF}, that implements the Landauer-B{\"u}ttiker approach\cite{But86} to quantum transport. 
This code allows us to restrict the electronic structure calculation to the valence electrons only, using ultrasoft
pseudo-potentials (US-PPs)\cite{PhysRevB.41.7892} to model the electron-ion interactions. 
Transport in heterostructures\cite{Geisler:15a} can be treated by taking a momentum component $k_x$ perpendicular to the transport direction into account and averaging the transmission probabilities,
\begin{equation}
T(E) = \sum_{k_x} w(k_x) T(E,k_x)
\label{eq:T_k}
\end{equation}
where $T(E,k_x)$ is the transmission probability of an electron with energy $E$ and perpendicular momentum $k_x$, and $w(k_x)$ is a weighting factor following from symmetry considerations. 
As before, the PBE generalized gradient approximation\cite{PhysRevLett.77.3865} of the exchange-correlation
functional was employed. 
The  electronic  wave  functions  were
expanded  in  a  plane-wave  basis  set  with  an  energy  cutoff
of 40~Ry. The vdW interactions were treated by the  Grimme-D2 method\cite{doi:10.1002/jcc.20495}. 
In contrast to the adsorption study, we consider a free-standing, single-layer WS$_2$ sheet as building block of an envisaged spintronics device. 
The electrical current flows along the zig-zag chains of WS$_2$, taken to define the $z$ direction, while the WS$_2$ sheet is assumed to be infinitely extended in the perpendicular $x$-direction. 
To establish electrical contact, the sheet is clamped between Zr electrodes touching the sheet both from above and below, see Fig.~\ref{fig:diode}. 
The supercell of the electrode region is constructed from a rectangular cell  of WS$_{2} $ with side length $\sqrt 3 a_{\rm WS_2}$ and $2a_{\rm WS_2}$ containing four W and 8 S atoms attached to Zr(0001) surfaces.
For the free-standing part of the WS$_2$ sheet as well as for the clamped part the lattice constant of WS$_{2} $ ($a_{\rm WS_2} =3.18${\AA}) was used; i.e. the lattice parameter of Zr in the electrode was adjusted to $a_{\rm WS_2}$. 
Periodic boundary conditions are assumed in the $x$ and $z$ directions for the calculations of the electronic ground state. The Brillouin zone of the leads and of the center region have been sampled using a 12$\times$1$\times$12  and 12$\times$1$\times$1 Monkhorst-Pack grid, respectively.
For geometry optimizations, the Hellmann-Feynman forces were required to be smaller than $10^{-3}$~Ry/bohr.

%%%%%%%%%%%%%%%%%%%%%%%%%%%%%%%%%%%%%%%%%%%%%%%%
\section{RESULTS AND DISCUSSION}

\subsection{Geometry and stability of WS$_{2}$-metal interface}

In order to determine  the stable location of WS$ _{2} $ on Co and Zr slabs, 
the binding energy ($E_b$) was calculated as 
\begin{align}
E_{b} = E_{\rm{M-WS_2}}- E_{\rm{M}}  - E^{\rm{strain}}_{\rm{WS_2}} \, , \qquad \rm{M=Co,Zr} \, ,
\end{align}
where $E_{\rm M}$ and $E_{\rm WS_2}$ are the total energies 
of  the M slab (cobalt, zirconium) and of the WS$_2$ sheet strained to the lattice constant appropriate for each case. 
We first determine the equilibrium geometry of  Zr/WS$ _{2}$ and then investigate the stable configuration of WS$ _{2}$  (in both rotational variants) on Co(0001).

\begin{figure}
 \includegraphics[scale=0.6]{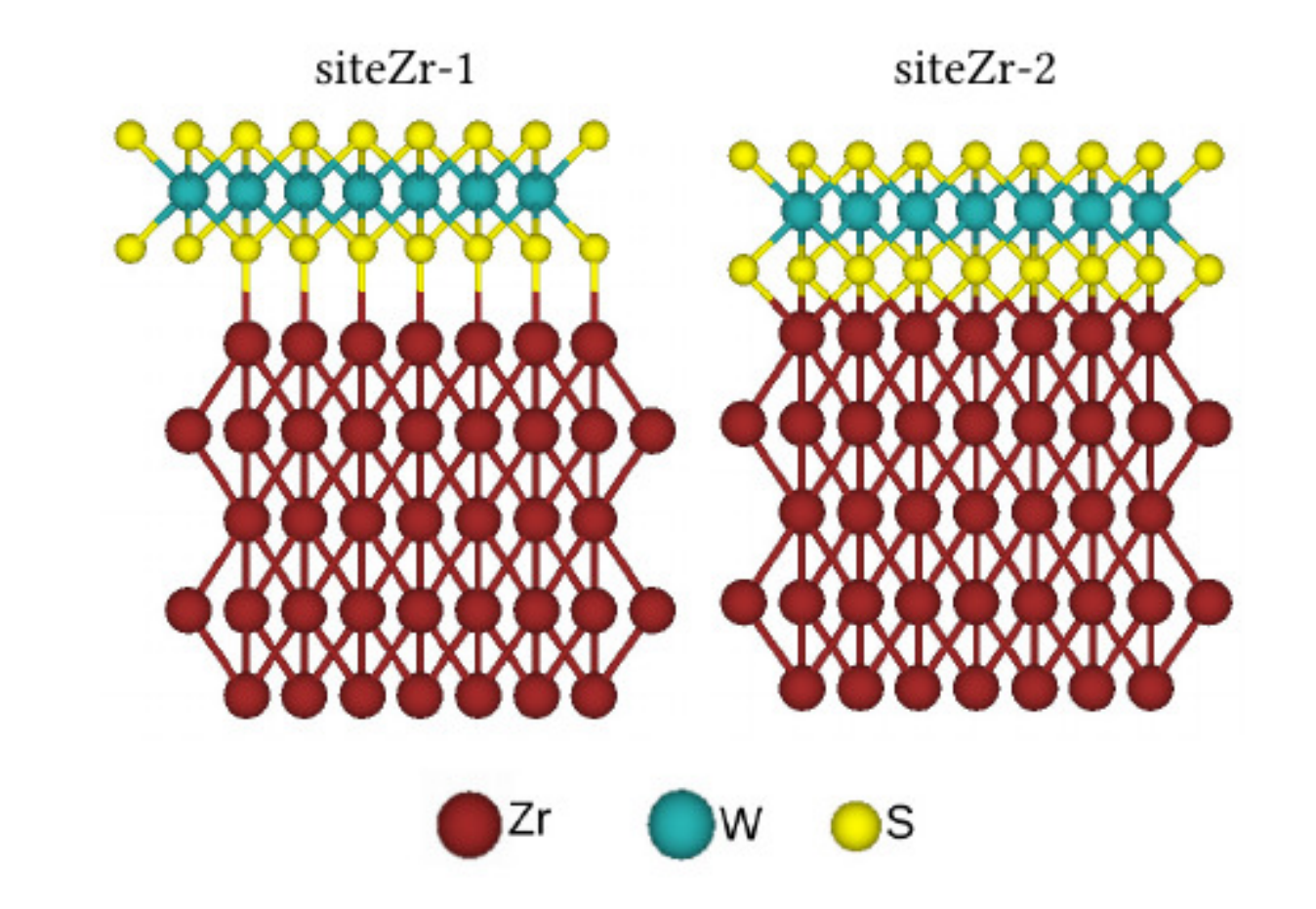}\\ 
\caption{\label{Zr}
 Atomic structure of Zr/WS$_{2} $  with two configurations.}
\end{figure}

\textbf{WS$_{2}$-Zr interface}:
 A five-layer Zr slab unit cell was used in the calculations, where the two bottom layers were fixed to the PBE bulk-optimized positions whereas the three top Zr layers were relaxed. We constructed two structures of Zr/WS$ _{2}$: 
 (i)   siteZr-1  configuration,  in which the W atom of the WS$_2$ layer sits above the  Zr atom of the second and forth Zr layer, and  the S layers are in line with the Zr atoms of  the first, third and fifth Zr layer. 
 (ii) siteZr-2 configuration, in which the W atom of the WS$_2$ layer is aligned with the Zr atoms of the first, third and fifth layer, and the S atoms are  located above the Zr atoms of the second and fourth Zr layer.  
 Optimized geometric structures are shown in Fig.~\ref{Zr}. We found that siteZr-2 is the most favorable configuration. 
The equilibrium distances $d_{\rm S-Zr}$  for siteZr-1 and siteZr-2  are  2.64 and  2.57{\AA}, respectively. These  interatomic distances are similar to the bond lengths which have been measured in the bulk zirconium sulfide (ZrS$ _{2} $) crystal with rocksalt structure (2.58{\AA}, Ref.~\onlinecite{ZrS}). 
The binding energies E$ _{b} $  follow the ordering  $1.92\mathrm{eV} = E_{b}$[siteZr-2] $ >  E_{b}$ [siteZr-1] $= 1.26$eV; i.e.,  
 smaller $d_{\rm S-Zr}$ correlates with a larger binding energy.

\textbf{Co/WS$_{2}$  interface}:
In the geometry optimization Co/WS$ _{2}R(0^{\circ})$ and Co/WS$ _{2}R(90^{\circ})$ , the WS$_2$ monolayer  
and the top cobalt layer were allowed to relax while the
remaining bottom layers were constrained to their bulk positions. 

\begin{figure}
\includegraphics[scale=0.5]{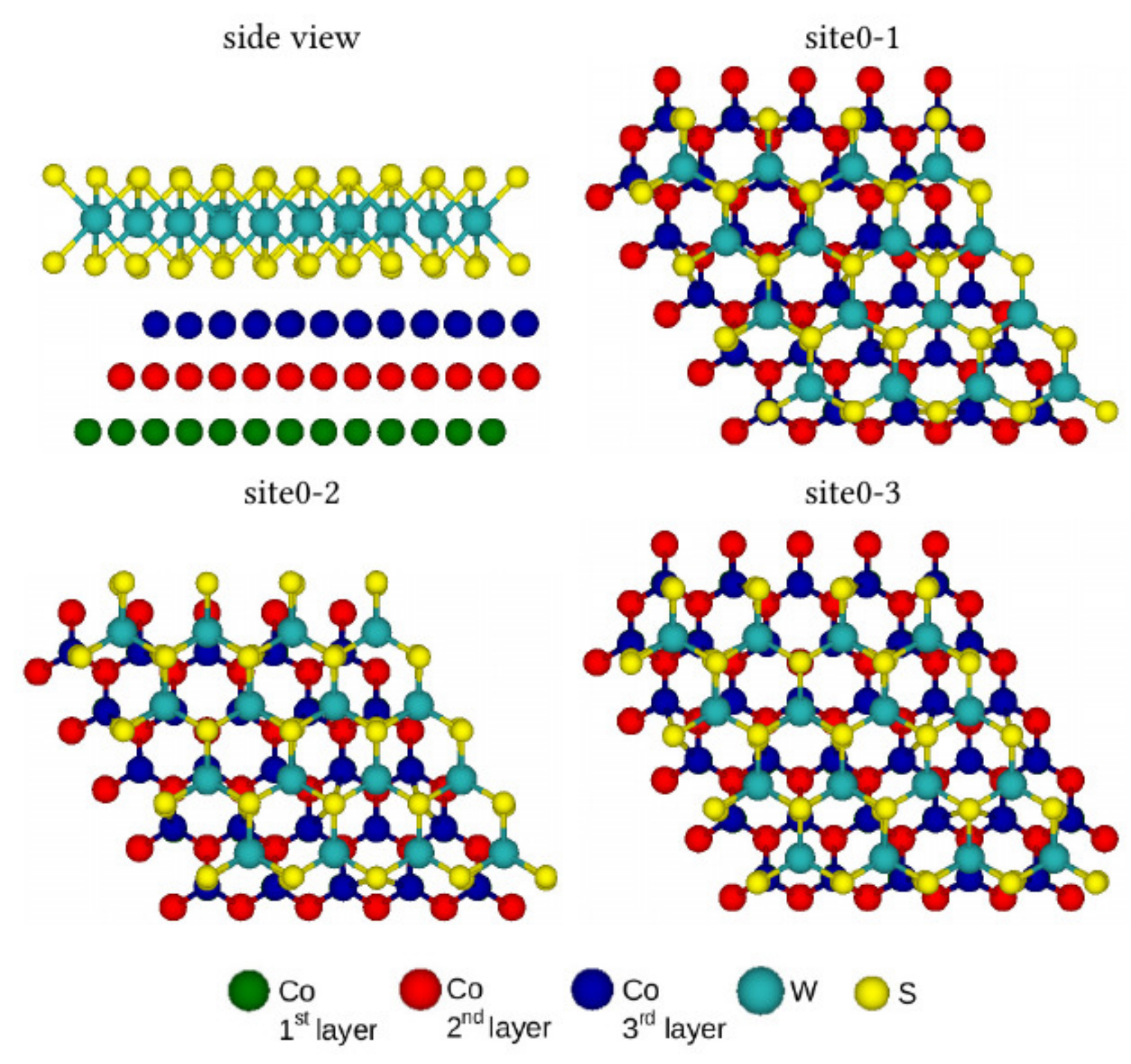} 
\caption{\label{str}
 Side and top views of monolayer WS$_{2}$ on the Co(0001) surface with parallel in-plane lattice vectors.}
\end{figure}

In order to study the structural and magnetic properties 
of  Co/WS$ _{2}R(0^{\circ})$,
we first looked for the stable configuration of  these heterostructures.
In this regard, we gradually move the WS$ _{2}$ parallel to the Co(0001) surface. 
Seven atomic configurations of the interface atoms were considered. 
After optimizing the structures from these seven initial configurations, we  found  that four  configurations transformed to  the site0-3 structure.
The structures found to be stable  are sketched in Fig.~\ref{str}.
The calculated binding energy as well as other structural parameters of 
the investigated configuration of Co/WS$ _{2}R(0^{\circ})$ are listed in Table~\ref{tab:str}.

\begin{table}
\caption{\label{tab:str} Relative total energies $ \triangle E $ (in meV),  
calculated binding energy $E_b$ (eV), 
average magnetic moment per cobalt atom M ($\mu_B$), 
average interface atomic distance ({\AA})  
of the three configurations of Co/WS$ _{2}R(0^{\circ})$ 
}
\begin{ruledtabular}
\begin{tabular}{lccccc}
&&Co/WS$ _{2}R(0^{\circ})$ &&\\
\hline
         &$ \triangle E $& $E_b$ &  M   & $d_{\rm Co-S} $     \\
\hline
site0-1 & 39.2 &-19.72 & 1.654 & 2.34 &   \\ 
site0-2 & 1.5& -19.76&  1.652 & 2.32 &  \\
 \textbf{site0-3} &\textbf{0.0}    & \textbf{-19.77} & \textbf{1.652}& \textbf{2.31} &  \\
\end{tabular}
\end{ruledtabular}
\end{table}

Our calculations show that site0-3 is the most stable configuration
of Co/WS$ _{2}R(0^{\circ})$. 
When the WS$ _{2} $ monolayer is  in contact to cobalt, the equilibrium distances $d_{\rm Co-S}$ 
range from 2.31 -- 2.34{\AA}. We can see that the smaller $d_{\rm Co-S}$ generally correlates with a larger binding energy and the site0-3 system has strong interface bonding. These  distances are  close to 
experimental data obtained for bulk  CoS$ _{2} $  with the pyrite structure (2.32{\AA}, Ref.~\onlinecite{exper}).

\begin{figure}
\includegraphics[scale=0.65]{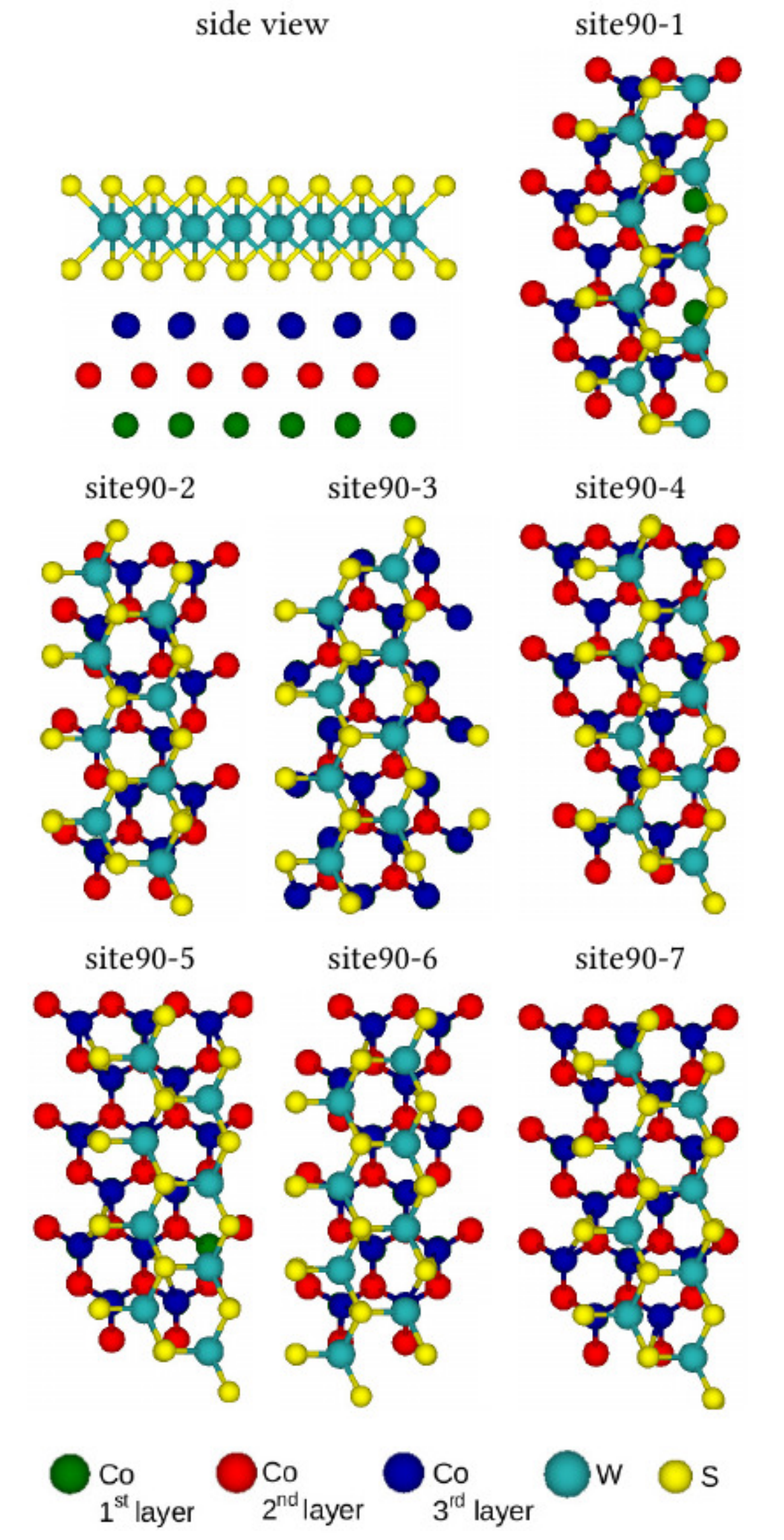}
\caption{\label{str90}
 Side and  top views of a $90^{\circ}$ rotated monolayer WS$ _{2} $ on the  Co(0001) surface.}
\end{figure}

To investigate the stable structure of Co/WS$_{2}R(90^{\circ})$, we moved WS$_{2} $ on the Co(0001) surface. 
All optimized geometric structures are shown in Fig.~\ref{str90}. We found that
site1-90 is the most favorable configuration (see Table~\ref{tab:str90}). The
differences in binding energy between  site1-90 and the other configurations are of the order of 
a few meV. 

\begin{table}
\caption{\label{tab:str90}
Relative total energies $ \triangle E $ (in meV),  
calculated binding energy $E_b$ (eV) per unit cell, 
average magnetic moment per cobalt atom M ($\mu_B$), 
average interface atomic distance ({\AA})   
of the different configurations of Co/WS$ _{2}R(90^{\circ})$ .
}
\begin{ruledtabular}
\begin{tabular}{lccccc}
&&Co/WS$ _{2}R(90^{\circ})$ &&\\
\hline
         &$ \triangle E $& $E_b$ &  M   & $d_{\rm Co-S}  $     \\
\hline
\textbf{site90-1} &\textbf{0.0} &\textbf{-5.146} &\textbf{ 1.636} & \textbf{2.342} &   \\ 
site90-2 & 0.021 & -5.145&  1.636 & 2.343 &  \\
 site90-3 & 4.792    & -5.141 & 1.635& 2.341 &  \\
 site90-4& 0.002   & -5.145 & 1.636& 2.343 &  \\
  site90-5 &  4.863 & -5.140 & 1.635&  2.341 & \\
    site90-6 & 4.884   & -5.140 & 1.635& 2.341 &  \\
     site90-7 & 0.181    & -5.144 & 1.636& 2.343 &  \\
   \end{tabular}
\end{ruledtabular}
\end{table}

\begin{figure}
\includegraphics[width=8cm]{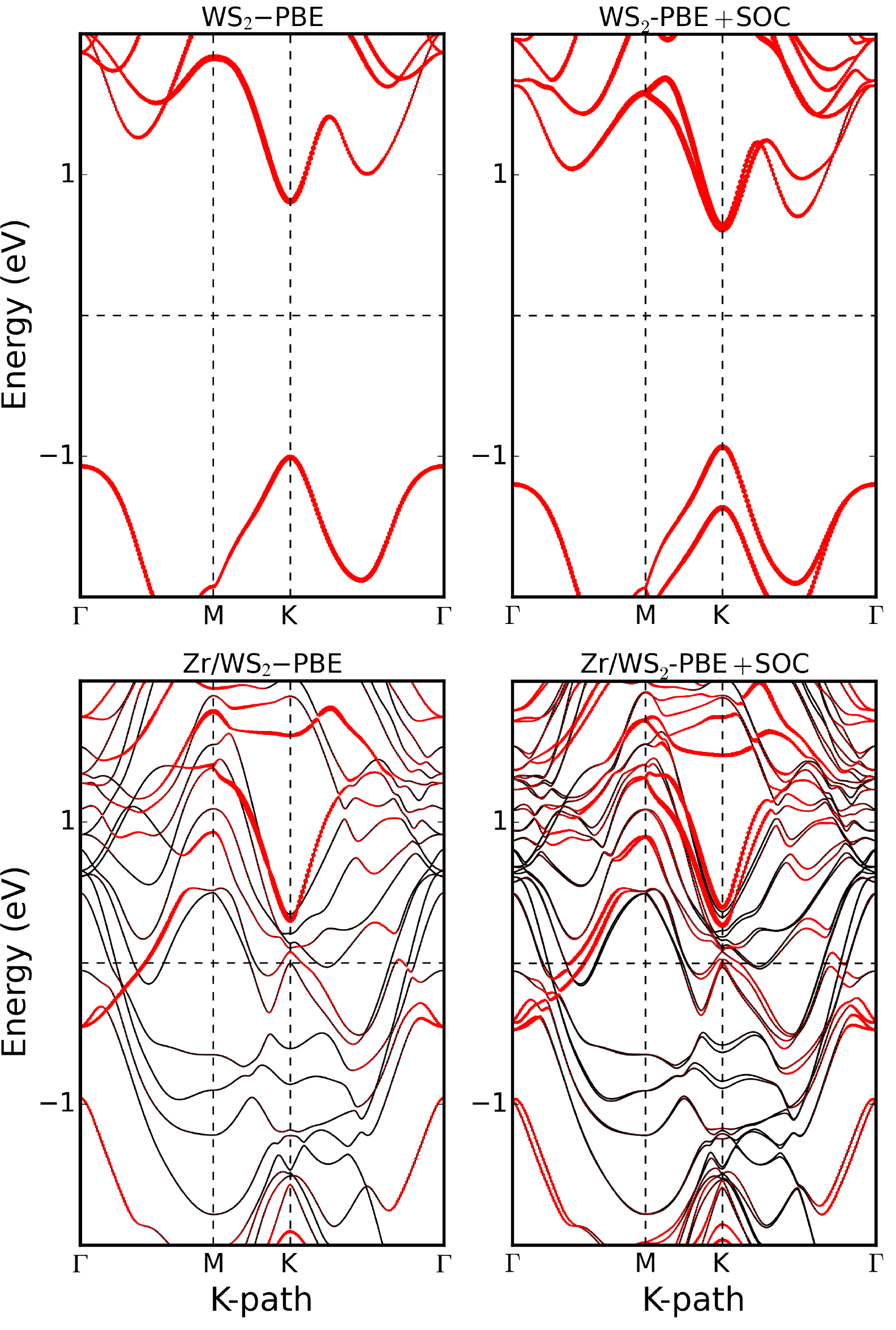}
%\begin{tabular}{lll}
%\includegraphics[scale=0.37]{fig5a.png}&\includegraphics[scale=0.37]{fig5b.png}\\ 
%\includegraphics[scale=0.37]{fig5c.png}&\includegraphics[scale=0.37]{fig5d.png}\\
%\end{tabular} 
\caption{\label{Fat-band}
Band structure and orbital character of single-layer WS$_2$ and Zr/WS$_2$ interface computed using the PBE and PBE+SOC methods. The thickness of the red lines  is proportional to the orbital weight of the W 5d orbitals in the respective band.}
\end{figure}

\begin{figure*}
\begin{center}
\includegraphics[width=14cm]{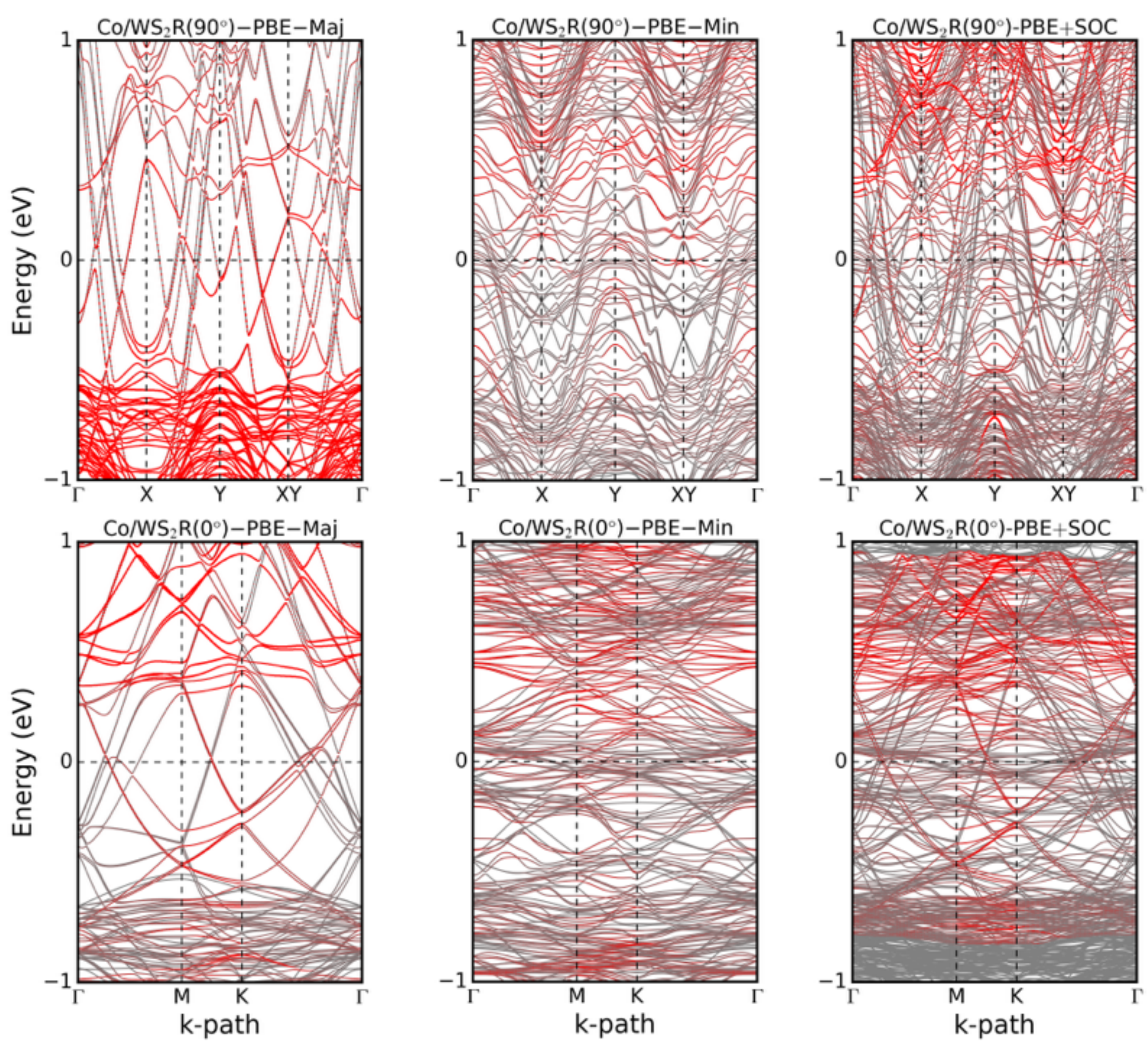}
\end{center}
\begin{tabular}{lll}
\end{tabular} 
\caption{\label{fatCo}
Band structure and orbital character of  Co/WS$ _{2}R(90^{\circ})$ (upper row) and Co/WS$ _{2}R(0^{\circ})$ (lower row) interfaces computed using the PBE and PBE+SOC methods, The labels Maj/Min indicate the majority-spin and minority-spin bands. The thickness of the red lines  is proportional to the orbital weight of the W 5d orbitals in the respective band.}
\end{figure*}
    
We  investigate the magnetic properties  of the most  stable of  both  Co/WS$_{2}R(0^{\circ})$ and Co/WS$_{2}R(90^{\circ})$. The results indicate that the average  local magnetic moment of a Co atom for Co/WS$ _{2}R(0^{\circ})$ (1.654$\mu_B$/Co)  and  Co/WS$ _{2}R(90^{\circ})$ (1.636$\mu_B$/Co) are  lower than in bulk Co (1.69$\mu_B$/Co) with the hcp structure.
To understand the magnetic moment of each atom in both structures, we considered the Mulliken population analysis  for these systems. 
In the interface Co layer, the magnetic moments are generally reduced compared to bulk Co and their values scatter between 1.45 and 1.67~$\mu_B$. 
The middle layer almost recovers the magnetic moments of bulk Co, while the Co atoms in the bottom layer of the slab with its free surface reach 1.75$\mu_B$/Co for  both Co/WS$_{2}$ slabs. The covalent bonding between interface Co and S atoms reduces the spin magnetic moment of the Co layer, while the interface S atoms become slightly magnetic (0.025$\mu_B$/S). 

Finally, to facilitate a direct comparison  among Zr/WS$ _{2}$, Co/WS$ _{2}R(0^{\circ})$  and Co/WS$ _{2}R(90^{\circ})$, 
the binding energy per interface atom was calculated for all three structures. 
In all three interfaces, the S atoms are covalently bonded to  Zr and Co, which means that the WS$_{2} $  single layer  is covalently bound to the Co  and Zr slabs and the van der Waals interaction plays only a minor role. 
The results show that the zirconium interface has a much stronger binding energy, $E_b = 1.92$~eV/Zr  atom,  with WS$ _{2}$ than cobalt. In Co/WS$ _{2} $, we find that the binding energy between WS$ _{2}$ oriented in the same way as the Co(0001) surface (0.79~eV/Co atom) is higher than for the Co/WS$_{2}R(90^{\circ})$ structure (0.43~eV/Co atom). 
This is because the alignment of the two hexagonal lattices of WS$_2$ and Co(0001) allows for better strain accommodation in the Co/WS$_{2}R(0^{\circ})$ interface.

\subsection{Electronic structure of WS$_{2}$-metal interfaces}

The orbital-projected electronic band structure of free-standing WS$_2$ 
and of the interfacial systems with Zr and Co are investigated  without (PBE) and with (PBE+SOC) 
spin-orbit coupling; and the results are presented in Fig.s~\ref{Fat-band} and \ref{fatCo}. 
Orbital-projected electronic bands  are displayed only for W atoms because the S, Zr and Co atoms are rather light and we therefore expect  their orbitals  not to be changed significantly under SOC. 
The free-standing monolayer WS$_{2} $ is a direct-gap semiconductor with a Kohn-Sham gap of 1.82~eV (1.54~eV when including SOC) located at the $K$ point of the Brillouin zone, in agreement with previous calculations \cite{Zibouche2014,2053-1583-4-1-015026,doi:10.1063/1.4774090}.
As indicated by the red color in Fig.~\ref{Fat-band},  both the valence band maximum (VBM) and the conduction band minimum (CBM) of WS$ _{2}$  are mainly composed of W 5d orbitals.
Taking into account spin-orbit coupling, we observe a large splitting of the VBM (420 meV) at the K point in the first Brillouin zone (see Fig.~\ref{Fat-band}). The splitting for the CBM is 30 meV and thus much  smaller than the VBM splitting, in agreement
with previous studies \cite{Zibouche2014,PhysRevB.94.115131}.
As can be observed in Fig.~\ref{Fat-band}, the band structure of WS$_{2}$ is strongly modified by the interface with Zr.
The semiconducting gap of WS$_{2}$ has disappeared and the Zr atoms introduce occupied bands into the gap which results in the Zr/WS$_{2}$  system displaying metallic behavior. 
In particular, an interface state along $\Gamma$ -- M with strong W 5d orbital character is formed that disperses upward and crosses the Fermi level. 
The CBM of WS$_2$ is still visible in the projection onto the W 5d orbitals as a parabolic band around the K point located $\sim0.4$eV above the Fermi level.  
By comparing the band structure obtained with and without SOC, one notices that the position of the Fermi level and the bands derived from Zr orbitals are almost unaffected. In the W-derived interface state and in the CBM at the $K$ point, a lifting of band degeneracy due to SOC is observed. 

To analyze the  band structure  for the Co contact, we performed a similar analysis for the Co/WS$ _{2}R(0^{\circ})$  and Co/WS$ _{2}R(90^{\circ})$  interfaces.
We observe that also the Co interface causes the semiconducting character of free-standing WS$ _{2} $ to disappear. New states cross the Fermi level, both for  the majority and the minority spin electrons (see Fig.~\ref{fatCo}), and consequently the system behaves like a  metal. 
In the majority spin band structures of both the Co/WS$ _{2}R(0^{\circ})$  and Co/WS$ _{2}R(90^{\circ})$  interfaces one can observe the fully occupied Co 3d states below $-0.6$ eV.  Some hybridization of these valence band states with the W 5d states is noticeable, but occurs far below the Fermi level.  
The contributions of the bands originating from WS$_2$ are most pronounced in 
the conduction band of the Co/WS$_{2}R(0^{\circ})$  and Co/WS$_{2}R(90^{\circ})$ interfaces and start at $\sim 0.3$ eV above $E_F$ in the majority spin channel. 
In the minority spin band structures of both the Co/WS$ _{2}R(0^{\circ})$  and Co/WS$ _{2}R(90^{\circ})$ 
a huge number of bands is observed near $E_F$. However, most of these bands are only weakly dispersive and originate from partially occupied Co 3d rather than from W-derived  orbitals. Electronic states with a mixed character start at $\sim 0.5$ eV above $E_F$. 
When SOC is included, the distinction between majority and minority spin states is no longer possible, but the combined band structure of both spins is very similar to a superposition of both spin band structures obtained previously without SOC. 
In the bands crossing $E_F$, a clear effect of SOC, e.g. SOC-induced avoided band crossings, is not observable.  
Also, the magnetic moments of the Co atoms change very little, by less than 1\%. 
 
\subsection{ Schottky barrier at WS$_{2}$-metal interfaces}

\begin{figure}
\begin{tabular}{ll}
 \includegraphics[scale=0.58]{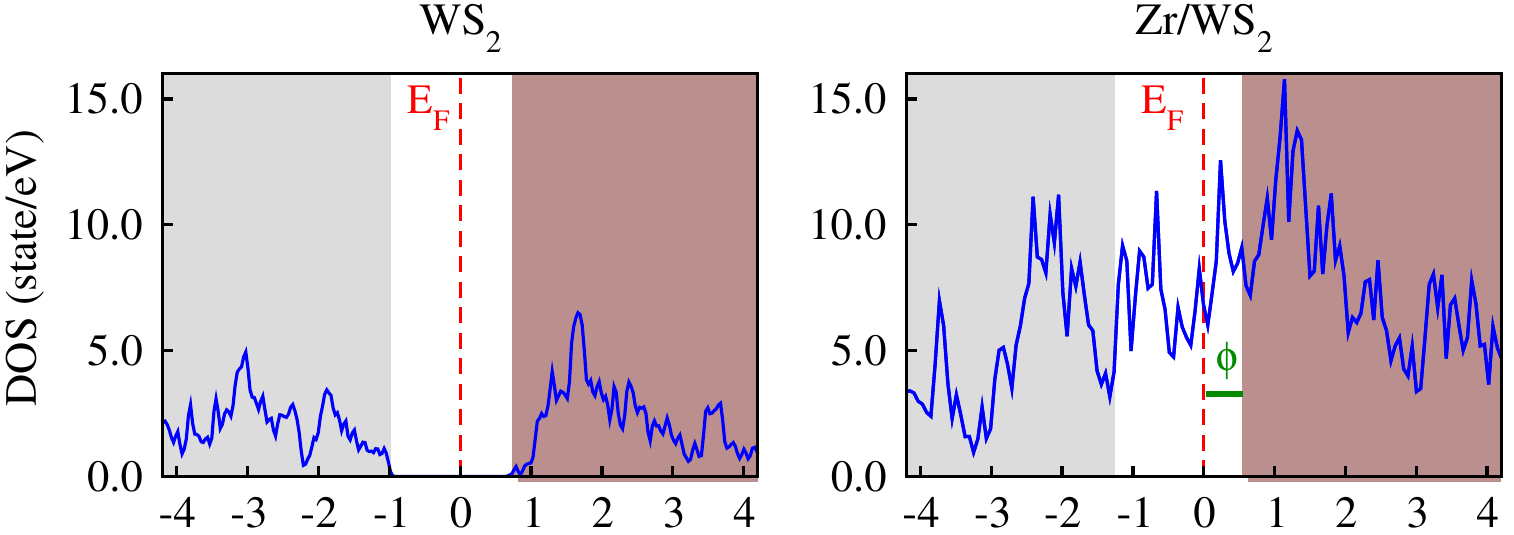} &\\
  \includegraphics[scale=0.58]{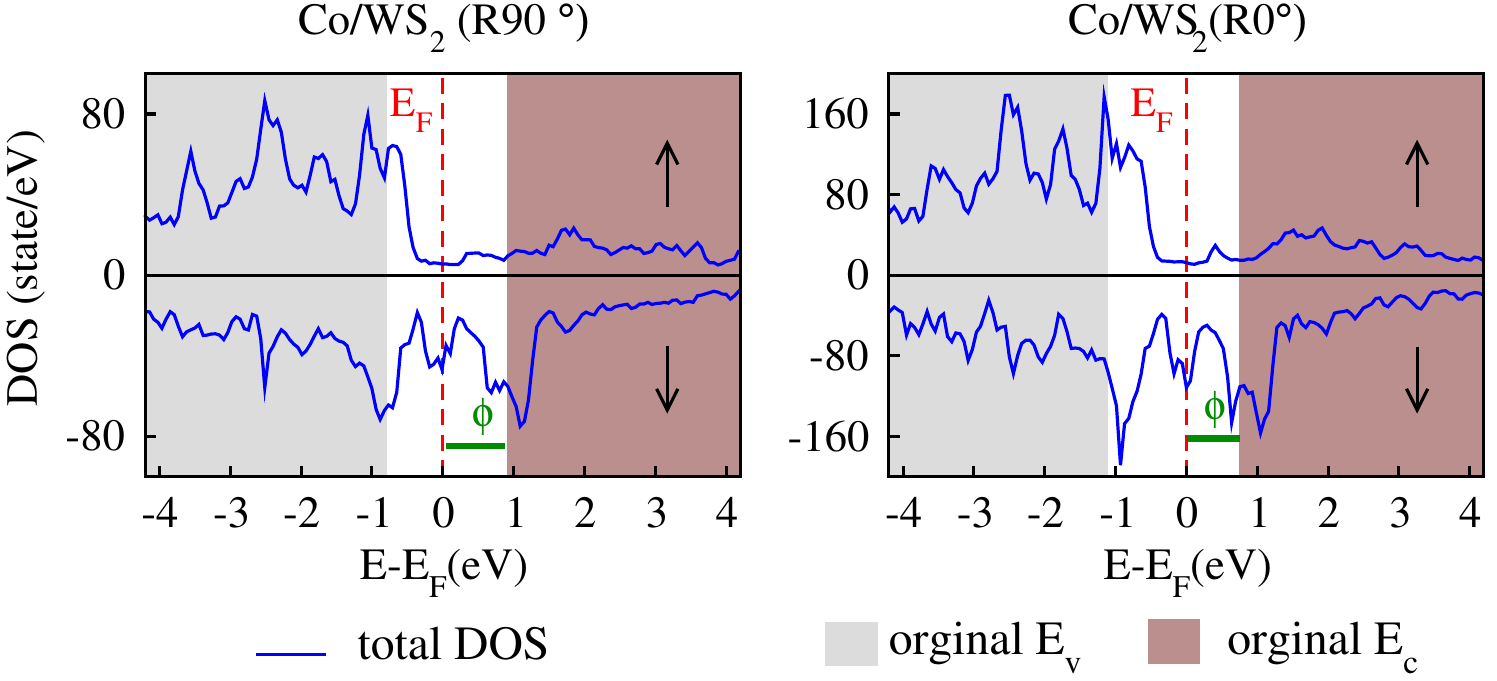} &
 \end{tabular} 
\caption{\label{SBS}
  Density of states for  WS$_{2}$, Zr/WS$_2$, Co/WS$_{2}R(90^{\circ})$ and  Co/WS$ _{2}R(0^{\circ})$ . 
  The energy regions of the conduction and valence band  of free-standing 
WS$_{2} $ is indicated by the brown and gray-shaded areas, respectively,
The Schottky barrier $\phi $ is marked by the green arrows. 
 The  Fermi level E$ _{F}$  is chosen as the zero of the energy scale. 
 The Fermi levels of the free-standing and the contacted systems have been adjusted using the W core levels as reference.}
\end{figure}

To evaluate the  Schottky barrier height (SBH) between the metal and the intrinsic (undoped) semiconductor WS$_2$, the  total DOS  of the M/WS$ _{2}$ (M=Zr,Co) interface displayed in Fig.~\ref{SBS} is compared to the DOS of single-layer, free-standing  WS$_{2}$.  
The Schottky barrier $\phi $   can  be  estimated from  the  energy  difference  between  $E_{F}$  of  the  interface system  and  the conduction band minimum (CBM) of WS$ _{2}$ located at the point $K$ in the Brillouin zone. 
The 1s core level of W is used as a common reference point for the Fermi levels both in the interface and in pure WS$_2$.  
As is seen from the figure, after interfacing WS$_{2} $ with Zr, the Fermi level is located above mid-gap, closer to the CBM, indicating the possibility to inject n-type carriers (electrons) into WS$_2$. 
Including the effect of spin-orbit coupling, the Zr contact to the WS$ _{2} $ monolayer shows a SBH of $\sim450$~meV.  
Furthermore, the  Co/WS$ _{2}R(0^{\circ})$ contact is found to be an n-type contact, too,  as seen in Fig.~\ref{SBS}. This is in line with recently reported   experiments and calculations for the Co contact with a MoS$_{2} $ monolayer \cite{PhysRevB.95.075402,doi:10.1021/nl4010157}. 
The  Schottky barriers are  $\phi  = 620$ and 
830~meV  for Co/WS$ _{2}R(0^{\circ})$  and Co/WS$ _{2}R(90^{\circ})$, respectively.    
When SOC is neglected, the electronic structure of the interfaces changes very little, but the gap of WS$_2$ and the position of $E_F$ in intrinsic WS$_2$ are affected (cf. Fig.~\ref{Fat-band}). As a result, the SBHs for all metal contacts are increased by 140 meV if SOC is neglected. 

\subsection{Quantum transport at WS$_{2}$  and Co/WS$_{2}$ }

Since the CBM and VBM of WS$_2$ are both located a the $K$ point, the WS$_2$ sheet in a nanoscale transistor 
should preferentially be oriented in such a way that the current flows along the $\Gamma-K$ direction in the Brillouin zone, i.e. along the atomic zig-zag chains of WS$_2$ in real space. 
The knowledge gained from the adsorption studies reported above are used to construct atomistic models of nanoscale spintronics devices. 
The most stable configurations of Zr/WS$_{2}$ and Co/WS$_{2}$ follow from the results of the previous section. 
 Although Co/WS$_{2}R(0^{\circ})$ is more stable than Co/WS$_{2}R(90^{\circ})$, we used the latter interface geometry to model a stripe of Co deposited on WS$_2$ because this leads to a computationally tractable set-up. A Co stripe with Co/WS$_{2}R(0^{\circ})$ would require a much larger supercell with more than 600 atoms, and such a system is too large for a quantum transport calculation even on a supercomputer. 
 For comparison, we also  consider  a smaller system, a Co cluster deposited on WS$_2$. 

The electronic conductance is calculated with the PWCOND code using an open
 system consisting of 
 (i) a scattering region comprising the WS$_{2}$  monolayer  and a chosen interface to the Zr/WS$ _{2} $ electrodes, and 
 (ii) the left and right (in principle semi-infinite) Zr/WS$_{2} $ leads. 
The scattering region has to be large enough to avoid the interaction between right and left leads. 
In our calculations, the length of this region is  36.70~{\AA}  along the transport direction (twelve unit cells between the electrodes) that included 24{~\AA} free-standing WS$ _{2} $  and 12.70~{\AA} Zr/WS$_{2}$, see Fig.~\ref{fig:diode}.
The leads are described by calculating the complex band structure of a periodic supercell containing four bottom layers and three top layers of Zr on WS$_{2}$.

 \begin{figure}\begin{tabular}{l}
 a)  \hfill \\
 \includegraphics[scale=0.65]{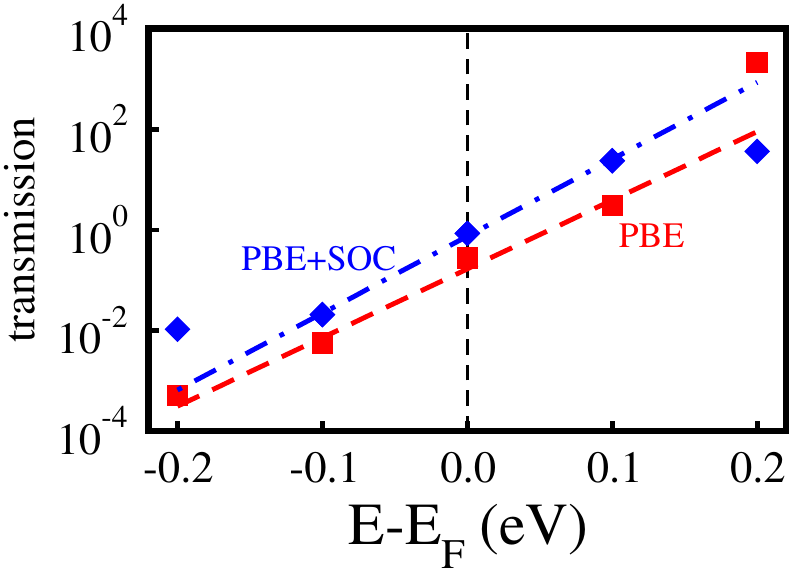} \\
 b)  \hfill \\
 \includegraphics[scale=0.65]{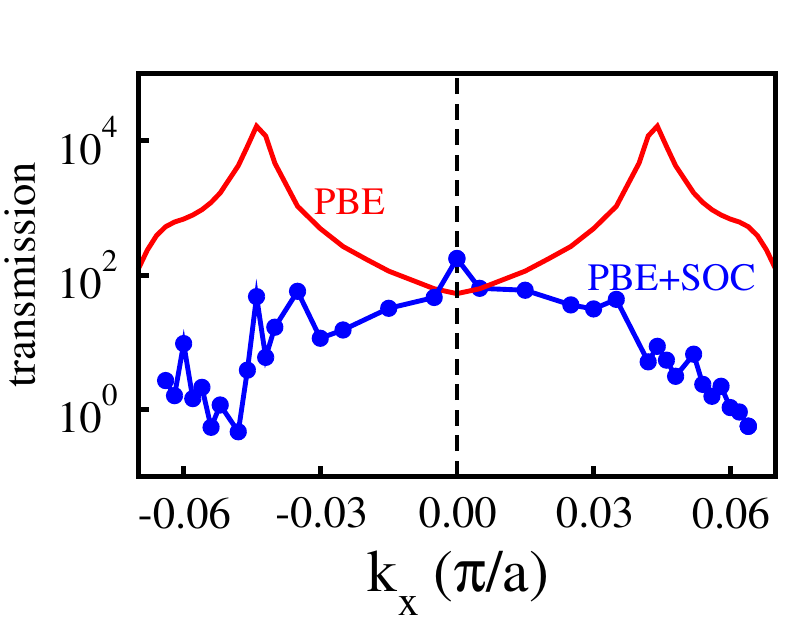}
  \end{tabular}
 \caption{\label{cond}
a) Transmission of the free-standing WS$_2$ sheet between Zr contacts, as shown in Fig.~\ref{fig:diode}, as a function of energy using the PBE (squares) and PBE+SOC (diamonds) methods, together with exponential fits (dashed lines). 
b) Transmission as a function of perpendicular momentum $k_x$ at $E_F+0.2$eV. }
 \end{figure}

\begin{figure*}
\begin{center}
\includegraphics[width=17cm]{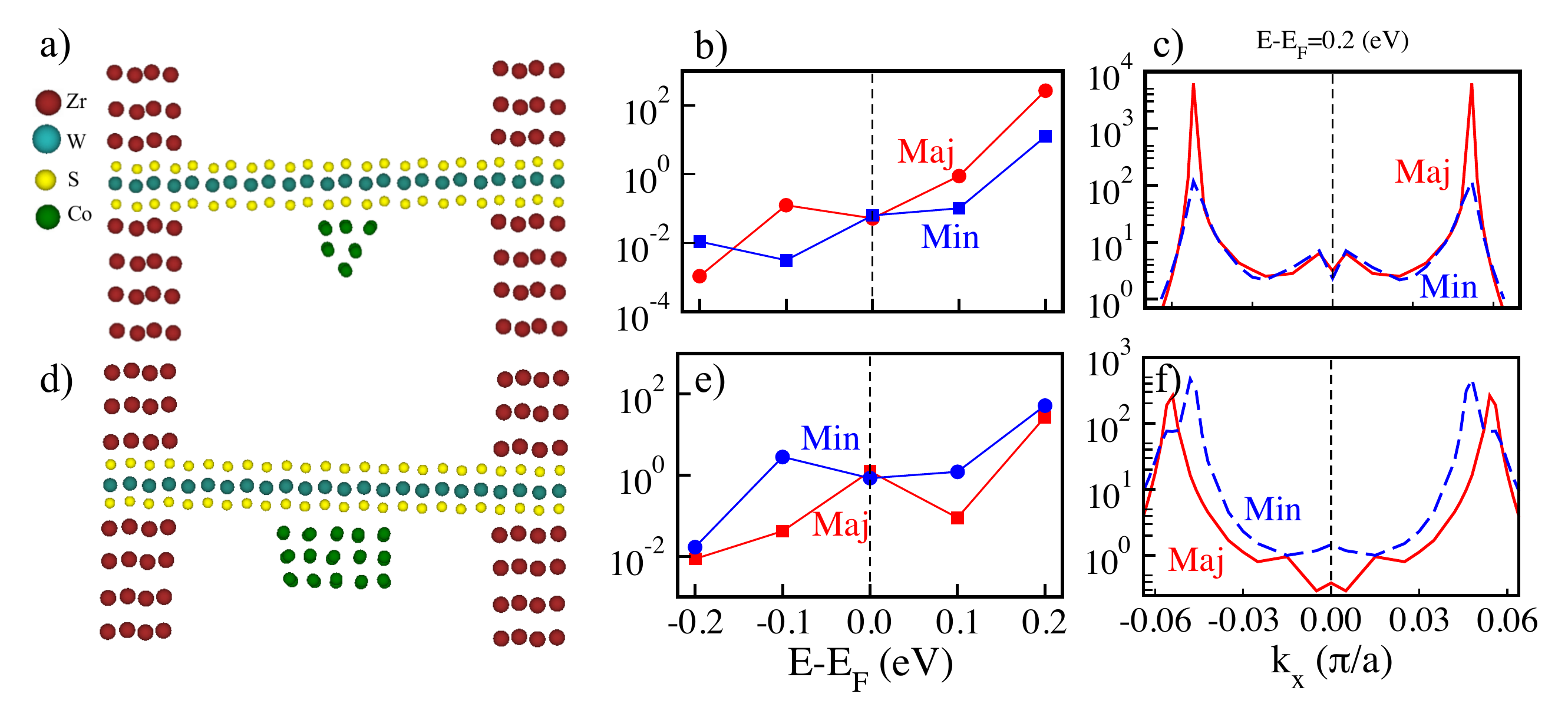}
\end{center}
%\begin{tabular}{ll}
%a) \hspace{2cm} WS$ _{2} $/Co cluster & b) \hspace{2cm} WS$ _{2} $/Co stripe\\
%\includegraphics[scale=0.4]{clusterco.pdf} & \includegraphics[scale=0.4]{chip.pdf} \\
%c) & d) \\
%\includegraphics[scale=0.85]{transCluster.pdf} & 
 %\includegraphics[scale=0.85]{translayer.pdf} \\
%e) & f) \\ \includegraphics[scale=0.85]{T-K-cluster.pdf} &
 %\includegraphics[scale=0.85]{T-K-layer.pdf} 
 %\end{tabular}
 \caption{\label{condCo}
 Optimized atomic structure (a,d), transmission as function of the electron energy (b,e), and transmission as function of perpendicular momentum $k_x$ at $E_F+0.2$eV (c,f) for the WS$_{2}$/Co cluster (upper row) and WS$_{2} $/Co stripe gate (lower row). }
   \end{figure*}

First we consider the quantum-mechanical transmission of a clean WS$_2$ sheet clamped between two Zr contacts. For injection into the conduction band of WS$_2$, the electrons must overcome a barrier (for macroscopic devices equal to the Schottky barrier calculated above). This is in line with the transmission $T(E)$ exponentially increasing with electron energy, as shown in Fig.~\ref{cond}a) both for the calculations with and without SOC.   
This device works like a diode with an I-V-characteristics described by $T^{-1} \, dT/dE = 35(eV)^{-1}$ for small electron energies. 
Following eq.~\ref{eq:T_k}, electrons with all possible momenta $k_x$ perpendicular to the transport  direction contribute to the overall transmission. 
With our choice of the supercell, the $K$ point is folded back to the $\bar \Gamma$ point of the supercell; thus we expect the main contribution from electrons with $k_x \approx 0$. This is confirmed by the results of the PBE+SOC calculations, see Fig.~\ref{cond}a). 
However, in the calculations neglecting SOC, a sharp rise of $T(E)$ is observed for electrons with an energy of 0.2eV above $E_F$. 
Analyzing the transmission as function of $k_x$, we find that this rise is due to maxima at $k_x = \pm 0.04 a_x/\pi$, see Fig.~\ref{cond}b).  We ascribe this observation to quantum-mechanical transmission resonances that depend both on the de Broglie wave length of the transmitted electron in the conductive channel as well as on the finite channel length. 
We note that, in the calculations neglecting SOC, states with wave vector $\vec k$ and $-\vec k$ must be equivalent due to time-reversal symmetry. 
In the calculations including SOC, transport with wave vector $\vec k$ and $-\vec k$ are no longer equivalent, but correspond to electrons at the inequivalent points $K$ and $K'$ in the Brillouin zone of WS$_2$. This also shows up in the asymmetry of the lower $T(E,k_x)$ curve in Fig.~\ref{cond}b).
 
Next we are adding a Co cluster 
or a  Co stripe  in the center region, as shown in Fig.~\ref{condCo}a) and d), and relax the geometry until we find a stable interface. 
The magnetization of Co breaks the time-reversal symmetry in the channel and mixes the $K$ and $K'$ valleys of WS$_2$. 
Because of this mixing and the qualitative similarity of the results for $T(E)$ with and without SOC, we consider it sufficient to study the 
transmission in the presence of a Co gate electrode without SOC. 
In Fig.~\ref{condCo}b) and e), we can see  plots of the spin-dependent transmission for
the WS$_{2}$/Co cluster and the WS$_{2}$/Co stripe. 
As an overall trend, the transmission curves are still increasing with energy, as in the case of pure WS$_2$, but show additional structure. 
The absolute value of the transmission is lower than in the clean case. 
Both observations point to a small additional barrier for the electrons imposed by the proximity of the Co cluster or stripe.  
This proximity effect is spin-dependent. 
For the WS$_{2} $/Co stripe the majority spin transmission is mostly (in particular for $E>E_F$) smaller than the minority spin transmission, while this trend is reversed for WS$_{2}$/Co cluster. 
We attribute these observations to quantum-mechanical interference effects which play a role in nanoscale devices such as the one considered here. 
The relation between a change in the channel potential and a change in transmission is, however, a subtle one, since 
not only the barrier height, but also the exact geometry and abruptness of the barrier affect the quantum-mechanical transmission amplitude. 
Analyzing the transmission as function of $k_x$ for the highest electron energy investigated, $E_F + 0.2$eV, we observe that the transmission resonance found for the pure WS$_2$ sheet has become very sharp by the addition of the Co gate, and displays different maximum transmission in the majority and the minority spin channel, see Fig.~\ref{condCo}c). The effect of a stripe-shaped Co gate is somewhat different: The resonances have nearly equal height in both spin channels, but their maxima as function of $k_x$ are slightly shifted, see Fig.~\ref{condCo}f). These equal heights explain why, nder 'open gate' condition at $E_F + 0.2$eV, both spin channels show nearly equal overall transmission and the spin polarization of the transmitted current tends to vanish.
 
\section{Conclusion}
In conclusion, we employed first-principles calculations to investigate various components of a future spintronics device based on a single-layer WS$_2$ sheet. Our calculations show that both Zr and Co can be used to form stable contacts on WS$_2$ that allow for injection of carriers into the conduction band. The most stable atomic structure of the interface can be deduced from the calculations, and it is shown that the magnetic moment of the Co surface persists in the presence of WS$_2$. Moreover, quantum-mechanical transport calculations show that a single-layer WS$_2$ sheet clamped between Zr electrodes behaves like a Schottky diode with a sharp I-V-characteristics. A gate electrode of Co can be used to control the spin polarization of the current. 
For the nanoscale device studied by us, we find that the spin polarization sensitively depends on the geometrical shape of the gate and on the energy of the electrons. As a direction for future work, one could attempt a generalization of these findings to larger devices where semiclassical transport theory would be applicable.  

\section*{Acknowledgements}
H.K. is grateful to the Iranian Ministry of Science, Research and Technology for a travel scholarship.  
We gratefully acknowledge the computing time granted by the Center for Computational Sciences and Simulation (CCSS) of the University of Duisburg-Essen and provided on the supercomputer magnitUDE (DFG Grant No. INST 20876/209-1 FUGG and INST 20876/243-1 FUGG) at the Zentrum f{\"u}r Informations-und Mediendienste (ZIM).

\end{document}